\documentclass[pre,aps,floats,superscriptaddress,floatfix,twocolumn,amsmath,amssymb]{revtex4}
\usepackage{amssymb,amsmath}
\usepackage{amsmath,amssymb}
\usepackage{graphicx}
\usepackage{psfrag}
\usepackage{multirow}
\usepackage{color}
\usepackage{dcolumn}
\usepackage{bm}
\usepackage{color}
\usepackage{enumitem,kantlipsum}
\usepackage{bm}
\usepackage[normalem]{ulem}

\def\beq{\begin{equation}}
\def\eeq{\end{equation}}
\def\bea{\begin{eqnarray}}
\def\eea{\end{eqnarray}}
\begin{document}
 \title{Kolmogorov or Bolgiano-Obukhov: Universal scaling in stably stratified turbulent fluids} 
 \author{Abhik Basu}\email{abhik.123@gmail.com, abhik.basu@saha.ac.in}
\affiliation{Condensed Matter Physics Division, Saha Institute of
Nuclear Physics, Calcutta 700064, West Bengal, India}
\author{Jayanta K Bhattacharjee}\email{jayanta.bhattacharjee@gmail.com}
\affiliation{School of Physical Sciences, 2A and 2B Raja S C Mullick 
Road, Calcutta 700032, West Bengal, India}

\date{\today}
\begin{abstract}


 We set up the scaling theory for stably stratified turbulent fluids. For a  
system having infinite extent in the horizontal directions, but with a finite 
width in the vertical direction, this theory predicts that the inertial range 
can display three possible scaling behaviour, which are essentially parametrised 
by the buoyancy frequency  $N$, or dimensionless horizontal Froude number $F_h$, and the vertical length scale $l_v$ that sets the scale of variation of the velocity field in the vertical direction, for a fixed Reynolds number. For very low 
$N$ or very high $Re_b$ or $F_h$, and with $l_v$ being of the same order as $l_h$, the typical horizontal length scale, buoyancy forces are irrelevant and hence, 
unsurprisingly, the kinetic energy spectra shows the well-known K41 scaling in the 
inertial range. In this regime, the local temperature behaves as a passively advected scalar, without any effect on the flow fields. For intermediate ranges of values of $N,\,F_h\sim {\cal O}(1)$, corresponding to moderate stratification, 
buoyancy forces are important enough to affect the scaling. This leads to the Bolgiano-Obukhov scaling which is isotropic, when $l_v\sim l_h$. Finally, for very large 
$N$ or equivalently for vanishingly small $F_h,\,L_o$, corresponding to 
strong stratification, together with a very small $l_v$, the system effectively two-dimensionalise; the kinetic 
energy spectrum is predicted to be anisotropic with only the horizontal part of the kinetic energy spectra follows the K41 
scaling, suggesting an intriguing {\em re-entrant} K41 scaling, as a function of stratification, for ${\bf 
v}_\perp$ in this regime. The scaling theory further predicts the scaling of the 
thermal energy in each of these three scaling regimes. Our theory can be tested 
in large scale simulations and appropriate laboratory-based   experiments.
 \end{abstract}
\maketitle
\section{Introduction}
Kolmogorov scaling arguments (hereafter K41)~\cite{kol,frisch} for the kinetic energy spectrum in inertial range, intermediate between the large forcing scales and small dissipation scales, of fully developed homogeneous and isotropic three-dimensional (3D) incompressible fluid turbulence provide the general basis for many theoretical studies on fluid turbulence. The K41 prediction, according to which the one-dimensional kinetic energy spectrum should scale as $k^{-5/3}$ in the inertial range, where $k$ is a Fourier wavevector belonging to the inertial range, has found good agreements with subsequent experimental and numerical studies~\cite{rahulrev}. The basic premise of the K41 arguments is that the inertial range scaling should depend only on the local wavevector $k$ and the constant (scale-independent) flux of the kinetic energy that is equal to the rate of the kinetic energy dissipation per unit mass.

When there is a density stratification, the resulting buoyancy forces can affect the inertial range scaling. Flows with buoyancy are important in geophysical setting, e.g., atmospheric wind flows and ocean currents. The buoyancy forces create temperature gradients leading to heat flows. In strongly and stably observed stratified flows, where the temperature gradient is positive (i.e., anti-parallel to gravity), quasi-horizontal layers have been observed in numerical simulations and laboratory-based experiments~\cite{exp1}. It is generally expected that for sufficiently small Ozmidov scale $L_o$ (see below for a formal definition), buoyancy forces should alter the inertial range scaling from the K41 prediction; see, e.g., Ref.~\cite{lind1}. Nonetheless, simulation studies of stratified turbulence reveal K41-like spectrum for the horizontal components of the velocity fields~\cite{riley1,almalkie}, a feature supported by geophysical measurements~\cite{lind1,lind2}. Such K41-like kinetic energy spectrum has been found in atmospheric turbulence as well~\cite{nastrom,cot}. Thus,  apparently the K41 scaling might hold even in situations with strong buoyancy, i.e., where it is {\em not} expected to hold. Several hypotheses have been developed to explain and clarify this apparently unexpected scaling behaviour. For instance, it has been attributed to an inverse cascade of energy, similar to conventional two-dimensional (2D) fluid turbulence~\cite{gage,lilly}. In contrast, simulations of strongly stratified turbulence in Ref.~\cite{herring} reported only a weak inverse cascade. Further, in case of stratified and rotating flows, inverse cascade has been observed only in the strong rotating limit, but not in the strongly stratified limit~\cite{metais}. On the other hand, Ref.~\cite{dewan} in fact suggested the existence of a forward cascade. It has been hypothesized in Refs.~\cite{lind2,lind3} of the existence of a strongly anisotropic three-dimensional (3D) in stratified flows with a forward cascade of energy. In yet another twist to this problem, several laboratory-based
studies using experimental tanks of different shapes and aspect ratios reveal an isotropic scaling for the energy spectra, that is {\em different} from the K41 scaling~\cite{bolg-lab}. This has been observed in the numerical studies of Refs.~\cite{rosen,mkv2,mkv1}.

The K41 scaling theory rests on the assumption of local isotropy, which is not maintained in stratified flows with sufficiently strong stratification~\cite{hebert}. Although this tentatively points towards the possibility of a scaling regime different from the K41 scaling, it should be kept in mind that distinctly anisotropic systems also can display the K41 scaling~\cite{yeung}. Recent numerical studies~\cite{kops1} however suggest that with increasing stratification, the inertial range scaling of the kinetic energy spectrum deviates from the K41 scaling. We thus see that the nature of universal scaling in stratified fluid turbulence - whether K41 or not - remains unsettled till the date. This prompted us to revisit the issue of scaling and its dependence on the buoyancy forces in turbulent stratified fluids by means of the scaling theory developed here. We first study the possible scaling regimes by re-writing the equations of motion in terms of dimensionless variables and show that there indeed three distinct scaling regimes, which are as follows. (i) Weak stratification when the effects of stratification is irrelevant, making buoyancy ineffective. The flow field is statistically identical to the conventional 3D fluid turbulence with the kinetic energy spectrum displaying the K41 spectrum, and the geometric anisotropy having no bearing on the the inertial range scaling. The temperature field is just passively advected by the ambient 3D velocity field. (ii) Moderate stratification: As stratification is increased, buoyancy becomes important, but anisotropy is not. The scaling of the kinetic energy spectrum is the well-known Bolgiano-Obukhov (BO) scaling, which is isotropic but  {\em different} from the K41 scaling~\cite{mkv2}. (iii) Strong stratification: For even higher stratification, both buoyancy and anisotropy are relevant, resulting into an anisotropic scaling by the kinetic energy spectrum, which now is dominated by the horizontal velocities. This makes it an effectively 2D turbulent system. The scaling of the thermal energy spectrum in each of these cases is also predicted. The rest of this article is organised as follows: In section~\ref{theory}, we revisit setting up the appropriate equations of motion for the velocity and temperature fields within the Boussinesq approximation. Then in section~\ref{scal-reg}, we heuristically argue the existence of three distinct scaling regimes, characterised by the strength of stratification and vertical length scales. Next, in section~\ref{scaling}, we set up the scaling theory to obtain the scaling exponents of the kinetic and thermal energy spectra in the different scaling regimes. Finally, in section~\ref{summ} we summarise and conclude our results.

\section{Equations of motion}\label{theory}

We consider a bulk stratified fluid that is infinitely extended in the   
horizontal directions, but has a finite extent of length $d$ along the vertical ($z$-) direction having an imposed density gradient. We consider a stable stratification, i.e., the imposed vertical density gradient is negative - the density (or, the local temperature) decreases monotonically as the vertical coordinate $z$ rises. 

We first revisit the derivation of the equations of motion. The Navier-Stokes equation for the 3D velocity field ${\bf v}= ({\bf v}_\perp,\,v_z)$, where ${\bf v}_\perp=(v_x,\,v_y)$ that include the buoyancy forces due to gravity is
\begin{eqnarray}
 \rho\left[\frac{\partial {\bf v}}{\partial t} + ({\bf v}\cdot {\boldsymbol\nabla}) {\bf v}\right] &=& - {\boldsymbol\nabla} p + \rho {\bf g} + \nu \nabla^2 {\bf v} + {\bf f},\label{basic-navier}
\end{eqnarray}
supplemented by the continuity equation for the density $\rho$:
\begin{equation}
 \frac{\partial \rho}{\partial t} + {\boldsymbol\nabla}\cdot (\rho {\bf v}) =0.\label{conteq}
\end{equation}
Here, $p$ is the pressure and $\nu$ a fluid viscosity~\cite{bulk-visc}, $\bf f$ is an external stirring force, required to maintain a nonequilibrium steady state, and ${\bf g}$ is acceleration due to gravity.

In the background, quiescent state ${\bf v}=0$ (no flow); however, there are pressure and density gradients along the $z$-direction. Let the total fluctuating pressure $p({\bf r}) = p_0 (z)  + \delta p({\bf r})$ and total density $\rho({\bf r}) = \sigma(z)  + \delta ({\bf r})$; $p_0(z),\,\sigma(z)$ are the background pressure and density fields in the quiescent state; $\sigma(z)$ describes the quiescent stratification.

The equation for the temperature fluctuations $\delta T$ is
\begin{equation}
 \frac{\partial\delta T}{\partial t} + {\boldsymbol\nabla}\cdot ({\bf v}\delta T)= \lambda\nabla^2 \delta T.\label{tempeq}
\end{equation}

In the Boussinesq
approximation~\cite{bouss} on (\ref{basic-navier}) that we now make, the density fluctuations are retained only in the 
buoyancy, and
neglected in every where else it appears in (\ref{basic-navier}). The buoyancy force is further expressed in terms of a fluctuating temperature assuming an equation of state of the form
\begin{equation}
 \rho=f(\delta T).
\end{equation}
In the Boussinesq approximation, (\ref{conteq}) reduces to just divergence-free condition on $\bf v$: ${\boldsymbol\nabla}\cdot {\bf v}=0$. Equation (\ref{basic-navier}) takes the form for ${\bf v}_\perp$ and $v_z$
\begin{eqnarray}
&&\frac{\partial {\bf v}_\perp}{\partial t} + ({\bf v}_\perp\cdot {\boldsymbol\nabla}_\perp) {\bf v}_\perp + v_z \frac{\partial {\bf v}_\perp}{\partial z} = - \frac{{\boldsymbol\nabla}_\perp\delta p}{\rho_0} + \nu\nabla^2 {\bf v}_\perp, \label{boussu}\\
&&\frac{\partial v_z}{\partial t} + ({\bf v}_\perp\cdot {\boldsymbol\nabla}_\perp) v_z + v_z\frac{\partial v_z}{\partial z} =-\frac{\partial \delta p}{\partial z} + \alpha g \delta T + \nu\nabla^2 v_z.\label{boussw}
\end{eqnarray}
Assuming an imposed temperature profile $\Delta T/d$, corresponding to a background imposed density profile varying linearly in the vertical direction, (\ref{tempeq}) reads
\begin{equation}
 \frac{\partial \delta T}{\partial t} + ({\bf v}_\perp\cdot {\boldsymbol\nabla}_\perp) \delta T + v_z\frac{\partial \delta T}{\partial z} = v_z\frac{\Delta T}{d} + \lambda\nabla^2 \delta T.\label{boussT}
\end{equation}
In the unforced, inviscid limit Eqs.~(\ref{boussu}-\ref{boussT}) admit the conservation of the total energy $E_{tot}$ defined as
\begin{equation}
 E_{tot}=\frac{1}{2}\int d^3x [ v_\perp^2 + v_z^2 + \frac{d}{\alpha g \Delta T} (\delta T)^2].\label{etot}
\end{equation}

\section{Scaling regimes}\label{scal-reg}

We first define Brunt-V\"ais\"al\"a frequency $N$ by
\begin{equation}
 N^2 = \frac{\alpha g\Delta T}{d}.\label{brunt}
\end{equation}

At this stage, it is convenient to introduce a horizontal scale $l_h$ and a horizontal velocity scale $U=(\epsilon l_h)^{1/3}$, where $\epsilon$ is the kinetic energy dissipation rate per unit mass. We further define a vertical scale $l_v$ that  characterises the vertical dependence of the velocity (treated as a free parameter here) and can be used to delineate different scaling regimes in the system (see below).  This allows us to define dimensionless velocities and temperature fluctuations by
\begin{eqnarray}
 &&{\bf u}=U {\bf v}_\perp,\, w= \frac{F_h^2U}{\mu}v_z,\, \delta T = \theta \Delta T,\nonumber \\&& \delta p = \rho_0 U^2,\, t=t\frac{l_h}{U},
\end{eqnarray}
where ${\bf u},\,w$ are the dimensionless horizontal and vertical velocities, and $\theta$ is the dimensionless temperature fluctuation; horizontal Froude number $F_h= U/(Nl_h)$ and $\mu$ is the aspect ratio $l_v/l_h$.   Then, $E_{tot}$ in (\ref{etot}) may be written as
\begin{equation}
 E_{tot}=\frac{U^2}{2}\int d^3x [u^2 + \frac{F_h^4}{\mu^2} w^2 + \frac{N^2d^2}{U^2} \theta^2].\label{etot1} 
\end{equation}
Furthermore, we define  $U=(\epsilon l_h)^{1/3}$, Reynolds number $Re=U^4/(\nu\epsilon)$, Ozmidov scale $L_o=\sqrt{\epsilon/N^3}$, buoyancy Reynolds number $Re_b=\epsilon/(\nu N^2)$ and dissipation scale $\eta=(\nu^3/\epsilon)^{1/4}$. Thence,
\begin{equation}
 \frac{N^2 d^2}{U^2}=\left(\frac{d}{L_o}\right)^{4/3}\left(\frac{d}{l_h}\right)^{2/3}.\label{factor}
\end{equation}
Now $L_o$ is the smallest scale up to which the effects of stratification or buoyancy will be significant. At scales smaller than $L_o$, practically no effects of buoyancy can be observed. Given that in pure fluid turbulence, the inertial range and the associated  K41 scaling persists up to the scale $\eta$ beyond which dissipation scale starts, if $L_o<\eta$, the whole of the scaling regime should be affected by buoyancy, where as, for $L_o>\eta$, inertial range scales bigger than $L_o$ are to be affected by buoyancy, but scales smaller than $L_o$ but bigger than $\eta$ should be unaffected by buoyancy. Hence a cross-over from a non-K41 type scaling regime to a K41 scaling regime is expected around scales similar to $L_o$.

In terms of the introduced dimensionless variables introduced above, Eqs.~(\ref{boussu}-\ref{boussT}) take the form
\begin{eqnarray}
 &&\frac{\partial {\bf u}}{\partial t} + ({\bf u}\cdot {\boldsymbol\nabla}_\perp) {\bf u} + \frac{F_h^2}{\mu} w\frac{\partial {\bf u}}{\partial z} = -{\boldsymbol\nabla}_\perp \delta p \nonumber \\ &+& \frac{1}{Re}\left(\nabla_\perp^2 {\bf u} + \frac{1}{\mu^2} \frac{\partial^2 {\bf u}}{\partial z^2}\right),\label{scaleu}\\
 &&F_h^2\left(\frac{\partial w}{\partial t} + ({\bf u}\cdot{\boldsymbol\nabla}_\perp)w + \frac{F_h^2}{\mu} w\frac{\partial w}{\partial z} \right) =\nonumber \\ && - \frac{\partial \delta p}{\partial z} + \frac{N^2 l_v^2}{\mu}\theta + \frac{F_h^2}{Re}\left(\nabla_\perp^2 w + \frac{1}{\mu^2} \frac{\partial^2 w}{\partial z^2}\right),\label{scalew}\\
 &&\frac{\partial\theta}{\partial t} + ({\bf u}\cdot{\boldsymbol\nabla}_\perp)\theta + \frac{F_h^2}{\mu}w\frac{\partial \theta}{\partial z} =\nonumber \\ &&\frac{w}{\mu} + \frac{1}{ReSc}\left( \nabla_\perp^2 \theta + \frac{1}{\mu^2}\frac{\partial^2\theta}{\partial z^2}\right),\label{scaletheta}
\end{eqnarray}
where $Sc=\nu/\lambda$ is the Schmidt number. 
While $d/l_h$ is a fixed number, $L_o$ is a measure of the strength of stratification, and varies; $L_o$ is the smallest scale up to which stratification or buoyancy is significant.  
We consider three different scaling regimes for large but fixed $Re$ (equivalently, fixed $\epsilon$):

(i) {\em Weak stratification:-} In this case, $N\rightarrow 0,\,L_o\rightarrow \infty,\,l_v/l_h\gg 1$, and both $F_h$ and $\mu$ become large with $F_h^2/\mu$ remains finite. With very low $N$ or very high $L_o$,  the left had side of (\ref{factor}) is very low, and the energy is dominated by the kinetic energy. In this case
\begin{equation}
  E_{tot}=\frac{U^2}{2}\int d^dx [u^2 + w^2]
 \end{equation}
predominantly.
 Further, 3D velocity $\bf v$ and temperature $\theta$ follow
\begin{eqnarray}
 \frac{\partial {\bf v}}{\partial t} + ({\bf v}\cdot {\boldsymbol\nabla}) {\bf v} &=& -\frac{\boldsymbol\nabla p}{\rho_0} + \nu\nabla^2 {\bf v},\\
 \frac{\partial\theta }{\partial t} + ({\bf v}\cdot {\boldsymbol\nabla}) \theta &=& D\nabla^2 \theta.
\end{eqnarray}
Clearly, in this limit the problem reduces to the passive scalar turbulence problem, where $\bf v$ evolve autonomously; since $l_v/l_h\sim {\cal O}(1)$ anisotropy is unimportant. Since $L_o$ diverges, it is larger than the inertial range scales, leaving the latter essentially unaffected by buoyancy. Since $F_h^2\sim \mu$, we find 
\begin{equation}
 \frac{l_v}{L_o}\left(\frac{l_h}{L_o}\right)^{1/3}\sim {\cal O}(1)
\end{equation}
corresponds to the region with K41 scaling in a state diagram in the $L_o-l_v$ plane (together with both $l_v$ and $L_o$ diverging). 

(ii) {\em Strong stratification:-} $N\rightarrow\infty,\,\,l_v/l_h\ll 1, \,F_h^2/\mu^2\rightarrow 0,\,\mu\rightarrow 0,\,L_o\rightarrow 0$. In this limit, the total energy reduces to
\begin{equation}
 E_{tot}=\frac{1}{2}\int d^3x N^2d^2\theta^2
\end{equation}
predominantly. Since $L_o$ is vanishingly small, inertial range scales are much bigger than $L_o$, and consequently are strongly affected by buoyancy. Further, focusing on the inertial ranges ($Re\rightarrow\infty$), (\ref{scalew}) gives way to
\begin{equation}
 -\partial_z\delta p + N^2 l_v l_h \theta = 0,
\end{equation}
implying no equation of motion for $w$, the vertical component of velocity. Furthermore, 
\begin{equation}
 \frac{\partial {\bf u}}{\partial t} + ({\bf u}\cdot {\boldsymbol\nabla}_\perp){\bf u} = -{\boldsymbol\nabla}_\perp \delta p + \frac{1}{Re}\left(\nabla_\perp^2 {\bf u} + \frac{1}{\mu^2}\frac{\partial^2 {\bf u}}{\partial z^2}\right).
\end{equation}
Thus, in the inertial range with $Re\rightarrow \infty$, $\bf v$ is effectively two-component ($w=0$), depending upon the 2D in-plane coordinates $x$ and $y$. Further, the incompressibility condition reduces to just ${\boldsymbol\nabla}_\perp\cdot {\bf u}=0$, implying a 2D incompressible flow field. It is then expected that ${\bf u}$ will display the K41 scaling, which is {\em re-entrant} in the $l_o-l_v$ plane as both $l_v,\,L_o$ reduce to vanishingly small.
Equation (\ref{scaletheta}) becomes
\begin{equation}
\frac{\partial\theta}{\partial t} + ({\bf u}\cdot {\boldsymbol\nabla}_\perp)\theta = \frac{w}{l_v} + \lambda(\nabla_\perp^2 +\frac{1}{\mu^2}\frac{\partial^2}{\partial z^2})\theta. 
\end{equation}
With $F_h^2/\mu^2\ll 1$ for strong stratification, we find
\begin{equation}
 l_v^2\gg L_o^{4/3}l_h^{2/3}
\end{equation}
corresponds to the region of 2D K41 scaling by the kinetic energy spectrum (together with the condition $l_v,\,L_o\rightarrow 0$). Since $L_o\ll \eta$, all scales belonging to the inertial range are affected by buoyancy.

(iii) {\em Moderate stratification:-} In this case $N\sim {\cal O}(1),\,\,l_v/l_h\lesssim 1,\,F_h^2/\mu\sim {\cal O}(1)$, giving $N^2d^2/U^2\gg 1$. Thence, in
\begin{equation}
 E_{tot}=\frac{U^2}{2}\int d^3x[u^2 + \frac{F_h^2}{\mu}w^2+ \frac{N^2d^2}{U^2}\theta^2],
\end{equation}
thus the kinetic and thermal contributions are of the same order.
Furthermore, since $\mu=l_v/l_h\sim {\cal O}(1)$, anisotropy should be irrelevant (in a scaling sense). All terms in each of (\ref{scaleu}-\ref{scaletheta}) are mutually competing. Since $L_o\sim \eta$ and $L_o/d< 1$, the inertial range scales are expected to be affected by buoyancy. Thus, we have a 3D system with a 3D flow field affected by buoyancy. We therefore expect an isotropic scaling different from K41 in the inertial range. Using $F_h^2/\mu\sim {\cal O}(1)$, we have
\begin{equation}
 l_v^2\sim L_o^{4/3}l_h^{2/3}
\end{equation}
describes the region corresponding to non-K41 type 3D isotropic scaling.

\begin{figure}[htb]
 \includegraphics[width=7cm]{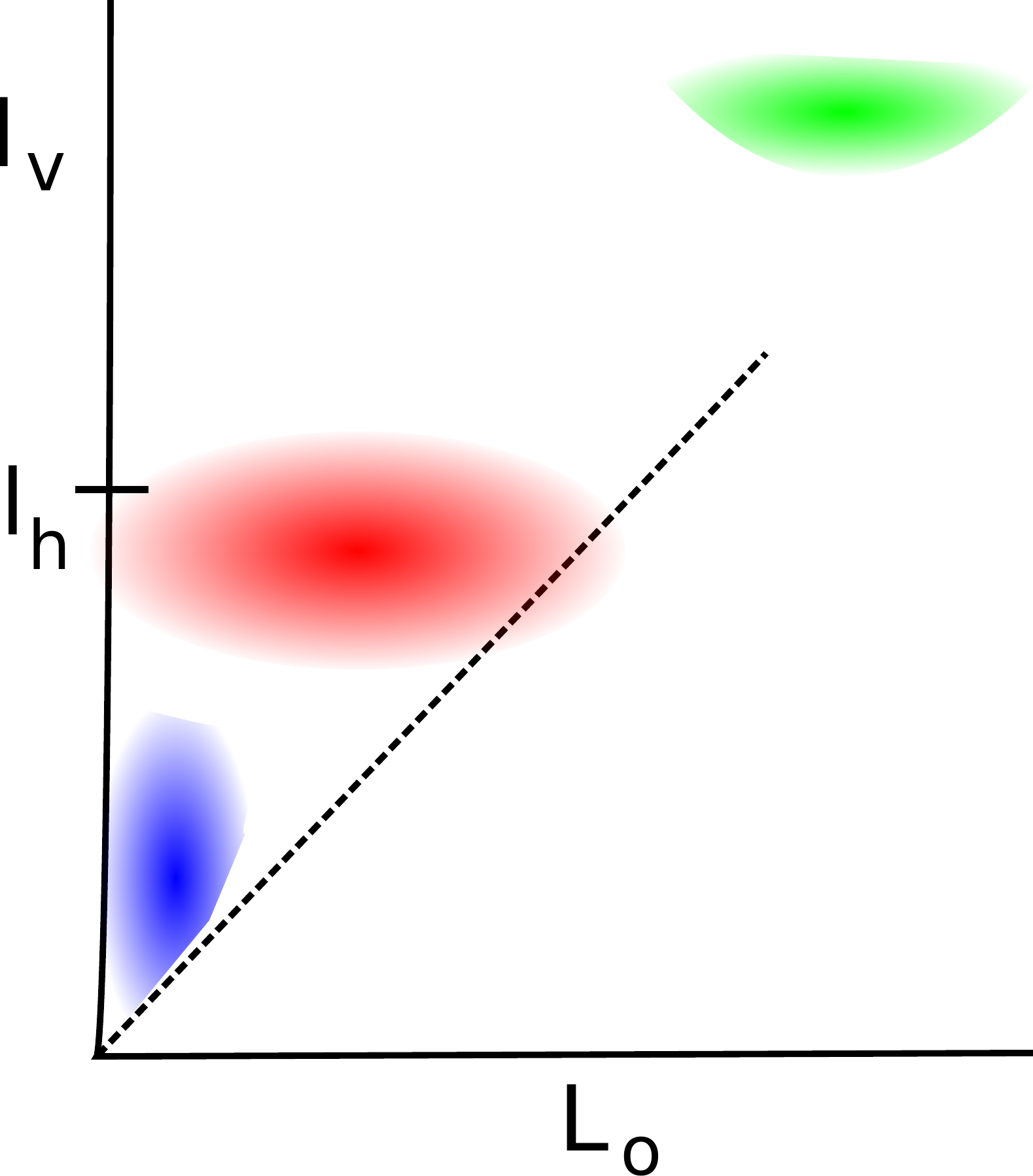}
 \caption{Schematic state diagram of stably stratified turbulence in the $L_o-l_v$ plane. The broken straight line indicates $l_v=L_o$. Three distinct scaling regimes are marked: the green patch at the top corresponds to weak stratification with 3D isotropic K41 scaling for the kinetic energy spectra, the red patch in the middle corresponds to moderate stratification with isotropic BO scaling for the kinetic energy spectra, and the blue patch corresponds to strong stratification with 2D K41 scaling for the in-plane part of the kinetic energy spectra. These regions are schematically drawn, have no sharp boundaries, and are surrounded by crossover regions (see text).}\label{state}
\end{figure}

A schematic state diagram in the $L_o-l_v$ plane is shown in Fig.~\ref{state}. The three regions correspond to the three different type of scaling are marked. Each region in Fig.~\ref{state} does not have any sharp boundary, and is surrounded by crossover regions.

We now extract the  scaling of the kinetic and thermal energy spectra by applying the scaling arguments originally formulated in Ref.~\cite{ab-jkb-mhd}.

\section{Scaling theory}\label{scaling}

In this Section, we closely follow Ref.~\cite{ab-jkb-mhd} and construct the scaling theory by using Eqs.~(\ref{boussu}), (\ref{boussw}) and (\ref{boussT}). To start with, it is instructive to  eliminate the pressure terms by imposing the incompressiblity condition. We obtain
\begin{eqnarray}
 \frac{\partial v_z}{\partial t} + P_{zj} ({\bf v}\cdot {\boldsymbol\nabla}_\perp){ v_z} &=& Ri\frac{\nabla_\perp^2}{\nabla^2}\theta + \nu \nabla^2 v_z + P_{zj}f_j,\label{eqz}\\
 \frac{\partial {\bf v}_\perp}{\partial t} + P_{\perp j}({\bf v}\cdot {\boldsymbol\nabla}_\perp){\bf v}_\perp &=& Ri\frac{{\boldsymbol\nabla}_\perp\partial_z}{\nabla^2} \theta + \nu \nabla^2 {\bf v}_\perp + P_{\perp j} f_j. \label{eqperp}
\end{eqnarray}
Here, ${\bf v}_\perp=(v_x,\,v_y)$, $P_{ij}=\delta_{ij}-\partial_i\partial_j/\nabla^2$ is the transverse projection operator, and $Ri$ is the Richardson number defined as $Ri=\alpha g d\Delta T$. The scaling theory to extract the scaling exponents for the kinetic and thermal energy spectra are built upon (\ref{eqz}), (\ref{eqperp}) and (\ref{boussT}).
 Decomposing three-dimensional (3D) wavevector ${\bf k} = ({\bf k}_\perp, k_z)$, we note that $k_{z-min}=2\pi/d$, due to the finite thickness along the vertical direction, while $k_\perp$ can be vanishingly small, since the system in unbounded along the horizontal directions. We mostly restrict ourselves to wavevector ranges with $k_\perp,\,k_z> 2\pi/d$, and briefly touch upon the scaling for $k_\perp \ll d$ at the end.

 We start by introducing the following rescaling of space, time and the fields:
\begin{eqnarray}
 &&{\bf r}_\perp\rightarrow l {\bf r}_\perp,\; z\rightarrow l^{\xi}\perp,\; v_z\rightarrow l^{a_\parallel} v_z,\; \nonumber \\ &&{\bf v}_\perp \rightarrow l^{a_\perp} {\bf v}_\perp,\; \theta \rightarrow l^y\theta,\;t\rightarrow l^{\tilde z} t. \label{basic-scal}
\end{eqnarray}
Notice that due to the anisotropic geometry of the system, we formally allow anisotropic spatial scaling as well as different scaling by $v_z$ and ${\bf v}_\perp$. Here, ${\bf r}= ({\bf r}_\perp,z)=(x,y,z)$. Further, the buoyancy force affects ${\bf v}_\perp$ and $v_z$ differently, as evident from Eqs.~(\ref{eqz}) and (\ref{eqperp}).  In addition, only $v_z$ appears linearly in (\ref{eqz}). We thus anticipate that for very strong buoyancy force, not only the scaling of the energy spectra should be anisotropic, even $v_\perp$ and $v_z$ should scale differently. This expectation is validated by the scaling theory below.

\subsection{Weak stratification: K41 limit}
In this limit, buoyancy forces are insignificant ($L_o\gg d$) and can be ignored from Eqs.~(\ref{eqz}) and (\ref{eqperp}). Then, the dynamics of $v_z$ and ${\bf v}_\perp$ become {\em autonomous}, independent of $\theta$. Then, Eqs.~(\ref{eqz}) and (\ref{eqperp}) are just the usual Navier-Stokes equation for an incompressible fluid, albeit for an anisotropic geometry. Thus when forced at large scales and in the infinite Reynolds number limit, ordinary 3D fluid turbulence characterised by the K41 kinetic energy spectra  should follow. This conclusion can be more formally obtained through the scaling theory.

We demand scale invariance of both (\ref{eqz}) and (\ref{eqperp}). This yields
\begin{equation}
 a_\perp = a_\parallel=1-\tilde z,\; \xi=1.
\end{equation}
Now impose scale independence of the kinetic energy flux. This yields~\cite{ab-jkb-mhd}
\begin{equation}
 2a_\perp = 2a_\parallel = \tilde z.
\end{equation}
Thus, we obtain $a_\parallel = a_\perp =1/3$ and $\tilde z=2/3$, corresponding to the K41 scaling by the kinetic energy spectra~\cite{ab-jkb-mhd}:
\begin{equation}
 E_v (k) \sim k^2 \langle |{\bf v}({\bf k},t)|^2\rangle \sim k^{-5/3}.
\end{equation}
Furthermore, demanding scale independence of the thermal energy flux, we find
\begin{equation}
 2y = \tilde z \implies y=\frac{1}{3}.
 \end{equation}
Thus, the thermal energy flux too displays K41 scaling:
\begin{equation}
 E_T(k)\sim k^2 \langle|\theta({\bf k},t)|^2\rangle \sim k^{-5/3}
\end{equation}
Interestingly, if we let the Richardson number $Ri$ to scale as $l^\eta$ then scale invariance of (\ref{eqz}) and (\ref{eqperp}) yields
\begin{equation}
 y+\eta = 1-2\tilde z = -\frac{1}{3} \implies \eta = -\frac{2}{3}.
\end{equation}
Thus $Ri$ scales down to zero in the long wavelength limit, recovering the ordinary fluid turbulence limit. This is consistent with the arguments made above.

\subsection{Moderate stratification: BO scaling}

Next we focus on the scaling when the buoyancy forces are  important ($L_o\lesssim d$).  Further, within this restricted zone of $k_\perp \sim k_z\gg 2\pi/d$ in the wavevector space, we expect the scaling should be isotropic to the leading order, and $v_z$ to scale in the same way as ${\bf v}_\perp$. We see below that these expectations are borne out by the scaling theory.

Demanding different advective nonlinearities in (\ref{eqz}) and (\ref{eqperp}) to scale in the same way, we find
\begin{equation}
 a_\perp = a_\parallel=a,\;\xi=1,
\end{equation}
as is expected. Next we balance the advective nonlinearities in (\ref{eqz}) and (\ref{eqperp}) with the buoyancy forces, as the latter is now relevant. We obtain
\begin{equation}
 ({\bf v}\cdot {\boldsymbol\nabla}){\bf v} \sim Ri \theta.
\end{equation}
Demanding that $Ri$ should be scale independent in a regime where the buoyancy forces are important and from the na\"ive scaling of the velocity (assuming no anomalous scaling), we find
\begin{equation}
 2a -1 = y,\; a=1-\tilde z.\label{scale-bo}
\end{equation}
Now imposing constant thermal energy flux in the inertial range, we find
\begin{equation}
 2y=\tilde z\implies a=\frac{3}{5},\; \tilde z=\frac{2}{5},\;y=\frac{1}{5}.\label{bol-scal}
\end{equation}
This gives
\begin{equation}
 E_v(k)\sim k^{-11/5},\;E_T(k) \sim k^{-7/5},
\end{equation}
as originally predicted by Bolgiano~\cite{bolg} and Obukhov~\cite{obu}.

 It would be interesting to see what might happen if we were to impose scale-independence of the kinetic energy flux, instead of the thermal energy flux. In this case, we would have $2a=\tilde z$ giving $a=1/3,\,\tilde z=2/3$. From (\ref{scale-bo}), we then find $y=2a-1=-1/3<0$! This appears unphysical, as $y<0$ implies $\theta$ becomes irrelevant (in a scaling sense) in the long wavelength limit. Nonetheless, noting that both $k_\perp,\,k_z$ are assumed to be larger than $2\pi/d$, it is possible to have a region just above $k_\perp,\,k_z=2\pi/d$ where this scaling holds, and hence cannot be just ruled out. The existence and implications of negative $y$ in intermittency that goes beyond simple scaling of the thermal energy spectrum would be interesting to investigate in experiments and numerical solutions of the equations. This is outside the scope of the present work.  

\subsection{Anisotropic scaling}

So far we have neglected the anisotropic effect while obtaining the BO scaling.  This cannot be generally true. For instance for $k_\perp \gg k_z$, the buoyancy forces clearly affect $\bf v$ anisotropically, i.e., affects $v_z$ and $v_\perp$ differently. Thus, scaling for $k_\perp \gg k_z>2\pi/d$ should be anisotropic, with $\xi$ to be either more or less than unity. As the buoyancy force rises, the isotropic region in the $k$-space shrinks, and the anisotropic region rises. This requires careful consideration.

To begin with we note that under rescaling as defined in (\ref{scaling})
\begin{equation}
 \nabla^2 = \nabla_\perp^2 + \partial_z^2 \rightarrow l^{-1}\nabla_\perp^2 + l^{-2\xi}\partial_z^2.
\end{equation}
We balance the linear terms in (\ref{eqz}), (\ref{eqperp}) and (\ref{boussT}). For $\xi >1$, we obtain from (\ref{eqz}) and (\ref{boussT})
\begin{equation}
 a_\parallel-\tilde z = y,\;\;y-\tilde z = a_\parallel.
\end{equation}
Thus $\tilde z=0$, which is unphysical. Hence, we discard the possibility $\xi >1$, and instead focus on $\xi <1$. 

In order to extract the scaling exponents, we balance the linear terms in (\ref{eqz}), (\ref{eqperp}) and (\ref{boussT}), impose scale independence of the thermal energy flux, and note that $a_\perp = 1-\tilde z$. We find
\begin{eqnarray}
 2\xi -1 + y &=& a_\parallel - \tilde z,\\
 y-\tilde z &=& a_\parallel,\\ 
 a_\perp - \tilde z = 1- 2\tilde z &=& \xi -1 + y,\\
 2y &=& \tilde z.
\end{eqnarray}
This gives 
\begin{equation}
a_\parallel = - y = -1/3,\,\tilde z=2/3,\,a_\perp = 1/3, \,\xi=1/3.\label{aniso-scaling}
\end{equation}
Interestingly, since $\xi <1$, and $a_\parallel < 0$, the system is effectively {\em two-dimensional} (2D)! In fact, with $a_\parallel<0$, $v_z$ strongly suppressed vis-\`a-vis ${\bf v}_\perp$. Further, with the scaling exponents given in (\ref{aniso-scaling}), we find for the in-plane component of the kinetic energy $E_{v_\perp}(k_\perp)$ that is constructed out of ${\bf v}_\perp$ only, and the thermal energy spectra $E_T(k_\perp)$ as functions of $k_\perp$ (see also Ref.~\cite{ab-jkb-mhd})
\begin{eqnarray}
 &&E_{v_\perp}(k_\perp) \sim k_\perp^{-5/3},\nonumber \\
 &&E_T(k_\perp)\sim k_\perp^{-5/3}, \label{aniso-spec}
\end{eqnarray}
indicating a {\em re-entrant} K41 scaling by $E_\perp(k_\perp),\,E_T(k_\perp)$, and hence by the total energy spectrum ( the contribution of $v_z$ to the kinetic energy spectrum is vanishingly small since $a_\parallel<0$). We thus find that the horizontal part of the kinetic energy spectra is {\em independent} of the stratification in the limit of very strong stratification, as suggested in~\cite{lind2}.  Note that this set of scaling exponents keep the kinetic energy flux, that now is dominated by ${\bf v}_\perp$, scale independent as well. Further, $\xi=1/3$ implies scaling $k_z\sim k_\perp^{1/3}$~\cite{naza}.

 It is interesting to consider the scaling of the damping terms in (\ref{eqperp}) and (\ref{boussT}) under the rescaling considered above. It is straightforward to see that in the isotropic scaling regimes (3D K41 or 3D BO), the viscosity $\nu$ and the thermal diffusivity $\lambda$ under the rescaling defined in (\ref{scaling}) with $\xi=1$ pick up scale factors $l^{2-\tilde z}$. With $\tilde z=2/3,\,2/5$, respectively, for 3D K41 and 3D BO scaling regimes, we find $\nu,\,\lambda\sim l^{4/3}$ and $\nu,\,\lambda\sim l^{8/5}$ in the 3D K41 and 3D BO scaling regimes, respectively. Extending this scaling argument to the strongly anisotropic case requires more care. First of all, the viscous and the diffusive terms in (\ref{eqperp}) and (\ref{boussT}) must now be generalised to appropriate anisotropic forms. Noting the rotational symmetry in the horizontal plane we  generalise the $\nu\nabla^2 {\bf v}_\perp$ in (\ref{eqperp}) to $(\nu_{zz}\partial^2_z +\nu_{\perp\perp}\nabla_\perp^2) {\bf v}_\perp$ and $\lambda\nabla^2 \delta T$ in (\ref{boussT}) to $(\lambda_{zz}\partial^2_z + \lambda_{\perp\perp}\nabla_\perp^2)\delta T$. Under rescaling (\ref{scaling}) with $\xi=1/3$, both $\nu_{zz}$ and $\lambda_{zz}$ pick up scale factors $l^{2y-\tilde z}=l^0$, where as both $\nu_{\perp\perp}$ and $\lambda_{\perp\perp}$ correspondingly pick up scale factors $l^{2-\tilde z}=l^{4/3}$. While the latter are same as for the usual 3D K41 scaling and progressively becomes large for larger spatial scales, $\nu_{zz}$ and $\lambda_{zz}$ do not scale at all, and hence do not undergo any singular renormalisation, an unexpected result from the scaling theory. In the large $l$ limit then $\nu_{\perp\perp}(l)\gg \nu_{zz}$ and $\lambda_{\perp\perp}(l)\gg \lambda_{zz}$. Whether this implies that damping is largely confined in the horizontal plane is a question that requires further study.

 Now formally allowing for a horizontal and vertical Richardson number $Ri_\perp$ and $Ri_z$, respectively, in place of $Ri$ in Eqs.~(\ref{eqperp}) and (\ref{eqz}), and demanding the scale invariance of (\ref{eqz}) and (\ref{eqperp}),
we find that $Ri_\perp$ is  scale-independent, where as $Ri_z$ decays as $l^{-2/3}$. Notice that the the anisotropic scaling exponents (\ref{aniso-scaling}) keep the horizontal part of the kinetic energy flux scale-independence, while the corresponding vertical part remains scale-dependent. This means $\tilde z=2/3$ {\em does not} control the renormalisation of the viscous coefficients in (\ref{eqz}). Allowing for a different dynamical exponents $\tilde z_\parallel$ for $v_z$ such that the vertical part of the kinetic energy flux is scale-independence, we find a value for $\tilde z_\parallel$, different from $\tilde z$, corresponding to weak dynamic scaling. In particular, $\tilde z_\parallel = 2a_\parallel =-1/3<0$, which is very unexpected! Accepting such a negative value of $\tilde z_\parallel$ apparently implies time-dependent correlation function of $v_z({\bf k},t)$ decays as $\exp (k^{-2/3}t)$ that decays very fast for small $k$ (we do not distinguish between $k_\perp$ and $k_z$ here). We are not aware of any measurements on time-dependent correlation functions of the velocity field in the presence of strong anisotropy; so this negative value of $\tilde z_\parallel$ remains purely speculative at present.

We can re-express the spectra (\ref{aniso-spec}) in terms of $k_z$ as follows. We note that
\begin{equation}
 E_{tot,a}=\int dk_\perp E_a (k_\perp) = \int dk_\parallel E_a (k_\parallel),
\end{equation}
where $tot$ refers to the total energy in the kinetic ($a=v_\perp$), or thermal ($a=T$) part. Hence dimensionally,
\begin{equation}
 E_a(k_\parallel)\sim \left[\frac{E_a(k_\perp)dk_\perp}{dk_\parallel}\right]\sim k_\perp^{-5/3}k_\perp k_z^{-1}.
\end{equation}
Now using $k_z\sim k_\perp^{1/3}$, we find the scaling
\begin{equation}
 E_a(k_z)\sim k_z^{-3},
\end{equation}
see, e.g., Refs.~\cite{naza,lumley,dewan1,hines,waite,lind2007}.

\subsection{Very low $k_\perp\ll 2\pi/d$}

 The scaling theory developed above holds for $k_\perp,\,k_z>2\pi/d$. What happens when $k_\perp \ll 2\pi/d$ (with $k_z\sim 2\pi/d$)? In this case, the buoyancy force are irrelevant in (\ref{eqz}) and (\ref{eqperp}). Further,  the incompressibility condition for wavevectors $k_\perp \ll 2\pi/d$ and $k_z\sim 2\pi/d$, reduces to (in the Fourier space) $k_zv_z (k_\perp\ll 2\pi/d,\,k_z\sim 2\pi/d)\approx 0$, giving $v_z (k_\perp\ll 2\pi/d,\,k_z\sim 2\pi/d)\approx 0$ to be ``small'': $v_z\sim (k_\perp/k_z) v_\perp$ in this wavevector range. Thus the flow field is to be dominated by ${\bf v}_\perp (k_\perp\ll 2\pi/d,\,k_z\sim 2\pi/d)$. Nonetheless, there is no 2D incompressibility on ${\bf v}_\perp$ in this wavevector range; the flow is still 3D incompressible. This, together with the irrelevance of the buoyancy force, leads us to speculate  that the scaling of the kinetic energy spectrum should be same as the K41 result, albeit with a strongly anisotropic amplitude.

\section{Summary and outlook}\label{summ}

We have set up the scaling theory for the inertial range scaling of the energy spectra in stably stratified turbulence. We show, by using a set of heuristic arguments and subsequently by our scaling theory, that for very weak stratification, the system is essentially identical to conventional fluid turbulence with both the kinetic and thermal energy spectra display the K41 scaling. For moderate stratification, scaling is expected to differ from the K41 prediction; nonetheless, anisotropy remains irrelevance, giving rise to a non-K41, but isotropic scaling, which we show to be the Bolgiano-Obukhov scaling. For very high stratification, anisotropy becomes relevant, leading to two-dimensionalisation of the system with the horizontal part of the kinetic energy spectra and the thermal energy spectra scale as per the K41 prediction. 

It is expected that the universal scaling properties
of fully developed stably stratified turbulence cannot be fully characterised by the two-point correlation functions (equivalently by the energy spectra) only. Instead, one needs to calculate a hierarchy of multiscaling exponents for different order structure functions (including
the two point ones) for the velocity and temperature fields. Our scaling theory is of course inadequate to obtain these multiscaling exponents. Nonetheless, it is reasonable to expect that the three different scaling regimes brought out by our scaling theory should actually correspond to three different multiscaling universality classes. Direct numerical simulations of the equations of motion should be useful for further studies in this regard. 

We have confined ourselves to the study of stably stratified turbulence here. Perhaps more common daily life example of unstable stratified turbulence is turbulence in a fluid that is heated from below. Such a system gets unstable once convection starts. It will be interesting to see whether and how our scaling theory can be applied to unstable stratified turbulence.

\section{Acknowledgement}
One of us (A.B.) thanks the Alexander von Humboldt Stiftung, 
Germany for 
partial 
financial support 
through the Research Group Linkage Programme (2016).


\begin{thebibliography}{99}
\bibitem{bulk-visc} Notice that we did not include any bulk viscosity term in (\ref{basic-navier}). This is in anticipation of an eventual incompressible fluid description within the Boussinesq approximation; see below.
\end{thebibliography}
\end{document}